\documentclass[multi]{cambridge6A}
\usepackage[numbers,sort,compress]{natbib}
\usepackage{rotating}
\usepackage{floatpag}
\rotfloatpagestyle{empty}
\usepackage{amsthm}
\usepackage{graphicx}

\begin{document}

\alphafootnotes
\author{Hua Chen and Allan H. MacDonald\footnotemark}
\chapter{Spin-Superfluidity and Spin-Current Mediated Non-Local Transport}

\footnotetext{Department of Physics, The University of Texas at Austin}

\begin{abstract}
Some strategies for reducing energy consumption in information processing devices 
involve the use of spin rather than charge to carry information. 
This idea is especially attractive when the spin current is a collective one 
carried by the condensate of a magnetically ordered state, rather than a quasiparticle
current carried by electrons or magnons.
In this Chapter we explain how easy-plane magnets can be viewed as Bose-Einstein condensates of 
magnons, defined in terms of quanta of the spin-component perpendicular to the easy plane, 
and how they can carry dissipationless spin-currents that induce  
non-local interactions between electrically isolated conducting channels.
We comment specifically on important differences between superconductivity  
in normal/superconducting/normal circuits and spin-superfluidity in 
normal/magnetic/normal circuits.  
\end{abstract}

\section{Introduction}

Spintronics,
the study of the interplay between the electrical transport
and magnetic properties of magnetically ordered 
solids, has made steady progress over the past few decades.
Spintronics involves both phenomena like giant magnetoresistance in which 
transport properties are influenced by magnetic order configurations, and phenomena like 
spin-transfer torques in which transport currents can be used to modify magnetic configurations.
Pure spin currents, which do not involve charge flow, are routinely detected via 
the spin-transfer torques they exert on magnetic condensates and the 
electrical signals they give rise to when spins accumulate near sample boundaries or at
electrodes. There are hopes that spin-currents have advantages over 
charge currents that can be exploited to enable faster or lower power electronic devices.
In this Chapter we discuss the notion of spin-superfluidity in thin film magnetic 
systems, either ferromagnetic or antiferromagnetic and either metallic or insulating,
that have approximate easy-plane magnetic order \cite{sonin_review, PhysRevLett.87.187202, PhysRevB.68.064406, Nogueira_spin_super, PhysRevLett.112.227201, chenPhysRevB.90.220401, AFMsuperPhysRevB.90.094408}.  
In spintronics spin-superfluidity
refers to the capacity for spin currents to be carried without dissipation
by a metastable configuration of a magnetic condensate, rather than by an 
electron or magnon quasiparticle current. 

Our Chapter is organized as follows. In Sec. 2 we introduce the 
concept of spin superfluidity using the common language of magnetism researchers by
applying Landau-Lifshitz equations to easy-plane magnets. 
To motivate the spin-superfluidity concept, 
we compare the spin-transport properties of easy-plane magnets to the
matter transport properties of an ideal classical fluid.  
At the end of the section we discuss some similarities and differences between easy-plane ferromagnets and BEC systems. 
In Sec. 3 we discuss perpendicular spin injection in finite-size easy-plane magnetic systems.
We then show that spin superfluids can exhibit Josephson-like IV characteristics that  
arise ultimately from the topological stability of easy-plane magnetic order in thin films. Finally, 
we discuss potential applications of this behavior, and also the influence in 
realistic materials of magneto-static interactions, magneto-crystalline anisotropy, 
and damped magnetization dynamics.
We conclude in Sec.4 with a discussion of the relationship between spin-superfluidity in 
easy-plane magnetic systems and superconductivity in metals.

This chapter is closely related to the chapter on Spintronics and 
Magnon Bose-Einstein Condensation by Duine, Brataas, Bender, and Tserkovnyak,
elsewhere in this volume. Both chapters are motivated 
by advances in spintronics that allow spin-currents to be routinely passed between 
different materials, including between metals and insulators.  The phenomena that are addressed 
in the Magnon Bose-Einstein Condensation chapter occur in easy-axis 
magnetic systems that are driven electrically into a quasi-equilibrium steady state.  

\section{Spin superfluidity in ideal easy-plane magnets}

To simplify the following discussion we represent an ideal easy-plane magnet by the Ginzberg-Landau free 
energy functional\cite{landau_statistical_book_95,PhysRevLett.87.187202} 
\begin{eqnarray}\label{eq:chen_Fxymag}
\mathcal{F} = \int dV \left [-|\alpha|{\bf M}\cdot {\bf M} + \frac{\beta}{2} ({\bf M} \cdot {\bf M})^2 + A|\nabla {\bf M}|^2 + K M_z^2 \right ].
\end{eqnarray}
In Eq.~\ref{eq:chen_Fxymag} the first two terms account for the magnetization magnitude and the ground state free energy.  
The third term is a magnetic stiffness energy $A|\nabla {\bf M}|^2 \equiv A (|\nabla M_x|^2 +|\nabla M_y|^2 +|\nabla M_z|^2) $ 
that parameterizes the free energy cost of magnetization non-uniformity. 
In the easy-plane case of interest, $K$ is positive and the last term characterizes the free energy cost of magnetization that 
is not oriented in the easy-plane. This expression ignores anisotropy within the easy plane, which we restore later,
and also the complex term with long-range non-locality
that accounts for magnetostatic interactions \cite{PhysRevLett.115.237201}. 
The dynamics of the magnetization $\bf M$ is described by the Landau-Lifshitz equation \cite{landau_statistical_book_95}
\begin{eqnarray}\label{eq:chen_LL}
\frac{d {\bf M}}{d t} = -\gamma {\bf M} \times \frac{\delta\mathcal{F}}{\delta {\bf M}},
\end{eqnarray}
where $\gamma = g\mu_B/\hbar$ is the gyromagnetic ratio and we have assumed $g$ to be negative for electrons. 
The Landau-Lifshitz equation are valid when the magnetization varies slowly in space and time and can be 
derived in a variety of different ways, for example starting from a density-functional theory of the magnetically ordered state \cite{PhysRevB.79.064403,PhysRevB.79.064404}. Using the free energy expression in Eq.~\ref{eq:chen_Fxymag} the effective magnetic field 
that appears on the right-hand side of the Landau-Lifshitz equations and drives magnetization precession is 
\begin{eqnarray}
\frac{\delta\mathcal{F}}{\delta{\bf M}} \equiv -{\bf H}_{\rm eff} = -2 A \nabla^2 {\bf M} + 2 K M_z \hat{z}.
\end{eqnarray}
It is sometimes stated that the Landau-Lifshitz equation is a classical equation which describes spin-angular momentum precession.
However, we prefer to view it as a quantum equation which describes the collective quantum dynamics of a 
magnetic order parameter; certainly its derivation is always quantum. It can be viewed as a classical equation only
because the quantum spin-dynamics of a macroscopic magnetic condensate is classical. In modern spintronics, the quantum nature of 
this equation is revealed by the appearance of $\hbar$ in the relationship between classical precession 
frequencies and spin electromotive forces \cite{emf-Berger,PhysRevLett.102.067201}.

The classical ground state of the easy plane ferromagnet has uniform in-plane magnetization. 
For small deviations from this classical ground state we parametrize $\bf M$ as $M_0(\cos\phi,\sin\phi,m_z)$ with $m_z = M_z/M_0 \ll 1$. In this limit, which we assume below, the Landau-Lifshitz equations take the form
\begin{eqnarray}\label{eq:chen_conjeom}
&&\dot{\phi} = 2\gamma K M_0 m_z,\\\nonumber
&&\dot{m_z} = 2\gamma A M_0 \nabla^2\phi.
\end{eqnarray}
(We have ignored terms that are higher order in the small quantities $\nabla\phi$ and $m_z$.) 
Note that the 2nd equation can be recognized as a continuity equation for $m_z$.  
In this interpretation the current corresponding to $m_z$ is the collective spin current
\begin{eqnarray}\label{eq:chen_jmacro}
{\bf j}_z = -2\gamma A M_0\nabla\phi.
\end{eqnarray}
The continuity equation is a direct consequence of the conservation of $m_z$ in an ideal easy-plane ferromagnet,
{\it i.e.} of the property that the Ginzburg-Landau energy is invariant under rotations around the $\hat{z}$-axis 
in magnetization space. As we discuss further below, a dissipationless spin current described by Eq.~\ref{eq:chen_jmacro} flows through the system when the system has nonzero $\nabla\phi$ \cite{PhysRevB.68.064406}. 

Sonin \cite{sonin_review} has proposed a helpful analogy between a magnetic system carrying a dissipationless spin-current 
and a rod that is twisted around its axis. The rod will rotate globally when a torque is 
applied at one end unless an opposite torque is applied at the other 
end. Although the net force on every individual atom in a twisted rod with balanced torques 
vanishes, the two torques can be viewed as giving rise to a uniform angular momentum flux,
an angular momentum supercurrent, which passes through the cross section of the rod and 
transmits a torque applied at one end to the other end, where it is compensated. 
The non-local relationship between remote ends of the rod is supported by the rigidity of the rod,
just as the non-local relationship between spin currents injected at opposite ends of an easy-plane magnet on 
which we focus is supported by the magnetic order parameter rigidity.  

It is important to observe that all the spin-supercurrent phenomena in equilibrium easy-plane magnets that
we comment on in this Chapter have an alternate description solely
in terms of the spin-torque language commonly used in spintronics, which applies 
to any magnetic system and is therefore more general.  
The analogy with superfluid phenomena is restricted to 
magnetic systems with easy-plane order, but is interesting nevertheless because 
of the properties it suggests, and because of the light it sheds on the relationship between the collective 
phenomena studied in superfluids and superconductors and those studied in modern magnetism research, in particular 
in spintronics. The conversion between normal metal currents and Cooper pair currents via Andreev scattering \cite{andreev,BTKPhysRevB.25.4515}, which is important in mesoscopic superconductivity, is simply the easy-plane 
limit of the spin-transfer torque concept so central in modern spintronics \cite{Slonczewski1996L1,Slonczewski1999L261,bergerPhysRevB.54.9353,bergerSTT, PhysRevLett.80.4281, Myers867, Sun1999157, Ralph20081190}. 
To better explain the relationship of spin superfluidity to other superfluid phenomena, 
we now briefly summarize some key properties of fluids and superfluids.

\subsection{Classical superfluids}

Part of the reason why easy-plane magnetic systems are usefully viewed as being {\it super} 
is that their properties are in compliance with the conventional definition of ideal fluids -- fluids without viscosity and thermal 
conductivity (adiabatic). 
Ideal fluids can be simply described by Newton's 2nd law, which is known in fluid dynamics as Euler's equation:
\begin{eqnarray}\label{eq:chen_eulereqn}
\frac{\partial \bf v}{\partial t} + {\bf v}\cdot \nabla{\bf v} = -\frac{1}{\rho}\nabla p,
\end{eqnarray}
where $\bf v$ is the velocity of an elemental volume of a fluid, $\rho$ is the density of the fluid, and $p$ is the pressure. Note that the left hand side is simply $d{\bf v}/dt$.

An ideal fluid has an important property, referred to as Kelvin's theorem \cite{landau_fluid_dynamics},
that the velocity circulation is time independent. (The velocity circulation is defined as the line integral of the velocity around any closed loop in the fluid.) 
We emphasize later that a related property is essential to the stability of supercurrent states in superfluids. 
For now we consider the case when the velocity circulation is zero, which means that the vorticity
\begin{eqnarray}
{\bf \omega} \equiv \nabla\times{\bf v}
\end{eqnarray}
vanishes identically everywhere in the fluid. Then one can always find a scalar function $\phi$ whose gradient is 
equal to the velocity, {\em i.e.} 
\begin{eqnarray}
{\bf v} = \nabla\phi.
\label{eq:jgradphi}
\end{eqnarray}
Eq.~\ref{eq:jgradphi} is similar to Eq.~\ref{eq:chen_jmacro}. Moreover, one can derive from Eq.~\ref{eq:chen_conjeom} an 
equation for $\bf j$ that looks similar to Eq.~\ref{eq:chen_eulereqn}, 
with the pressure term in the latter replaced by a term proportional to $m_z$. ( We will return to this point in the next subsection.)
Thus an easy plane ferromagnet can indeed be viewed as an ideal fluid with density proportional to the perpendicular component of the magnetization.
 
In the following we focus our attention on thin film systems in which 
the magnetization direction depends only on two spatial coordinates, since this is normally the 
case of greatest practical interest. The analogies we make will therefore be between
thin film magnets and two-dimensional fluids. What is different between an easy-plane-magnet quantum superfluids 
and the classical ideal fluid is that the velocity potential $\phi$ is identified as a phase or azimuthal orientation angle in the quantum case.
The line integral of the phase or angle gradient over any closed loop
must then be an integer multiple of $2\pi$. This circulation quantization leads to vortices, 
topological defects carrying nonzero circulation quanta. Since circulation is conserved in the bulk of an ideal fluid, a vortex will remain stable unless it reaches the boundary of the fluid where circulation is not well defined, or it annihilates with another vortex with opposite circulation. In circular coordinates $(r,\theta)$ a vortex with 
circulation $\kappa$ can be represented by the velocity field
\begin{eqnarray}
{\bf v} = \frac{\kappa}{2\pi} \frac{\hat{\theta}}{r}.
\end{eqnarray}
One can then estimate the kinetic energy associated with a vortex by integrating $\frac{1}{2}\rho v^2$ over the whole fluid.
It is easy to see that the energy of the vortex increases logarithmically with the system size. 
It follows that creation or annihilation of a vortex is associated with an unbounded energy change. 
Moreover, under the assumption of zero viscosity and adiabaticity, creation and annihilation of vortices is the only way for a superfluid to relax from a metastable state with nonzero supercurrent to the zero current ground state. 
Because the creation of these topological defects requires that large energy barriers be overcome, the current state of a 
superfluid is extraordinarily stable.

\begin{figure}
\includegraphics[width=6.5cm]{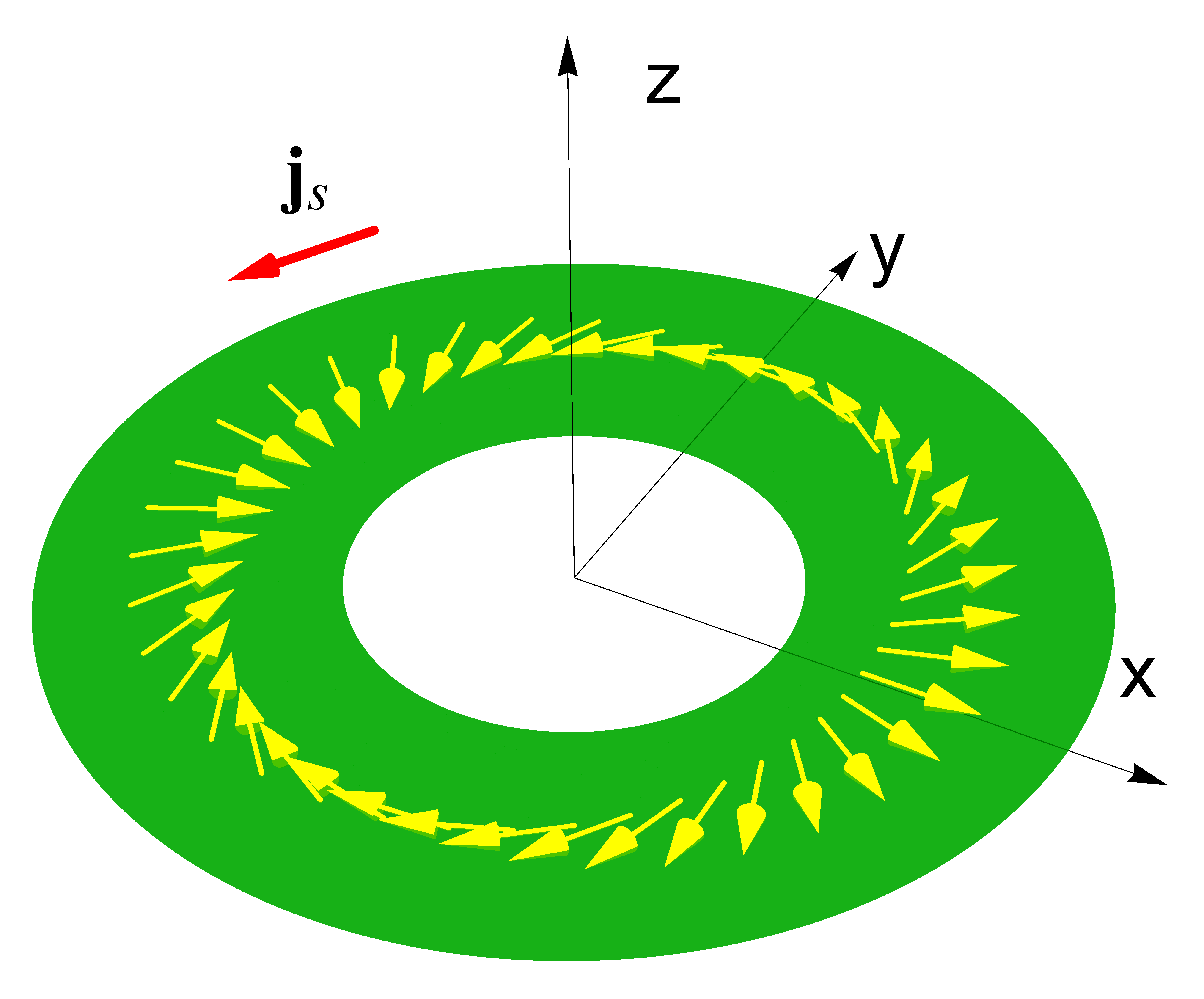} 
\caption{Metastable magnetization configuration formed by an easy-plane ferromagnet in a ring. This configuration carries a dissipationless spin supercurrent. The magnetization in this illustration changes by $4\pi$ upon enclosing the ring.}
\label{fig:chen_fig1}
\end{figure}

It is instructive to consider the example in which we connect the two ends of a long thin ferromagnetic wire to form a ring, 
as shown in Fig.~\ref{fig:chen_fig1}. The in-plane magnetization angle must then rotate by an integer multiple of $2\pi$ as one moves around the ring to complete a cycle. Provided that the total rotation angle is not zero, there is according to Eq.~\ref{eq:chen_jmacro} a persistent spin supercurrent in the ring because of the nonzero azimuthal angle gradient. 
The topological stability of this spin supercurrent state is then obvious since it is not possible to change the 
angle winding number by locally perturbing the magnetization. This geometry is similar to the twisted rod example given earlier in this chapter, and is related to the celebrated rotating cylinder experiment in superfluid He$^4$. 

Now imagine that a vortex with an angle winding of $2\pi$ is nucleated at one boundary of the ring and moves across the width of the ring. 
The azimuthal angle change from one end of the sample to the other, measured along the direction perpendicular to the path of the vortex, changes by $2 \pi$ for every vortex which is nucleated on one edge of the ring, moves across, and is then
annihilated at the other edge to restore a uniform superfluid. This {\it phase slip} is accompanied by a lower free energy when it reduces $|\nabla{\bf M}|^2$, and also a smaller spin supercurrent. 
The barrier for supercurrent relaxation is thus proportional to the vortex nucleation energy, and can greatly exceed $k_B T$ because it is a collective barrier involving many electronic spins.
A similar argument explains the metastability of currents in superconductors. In the interior of a magnetic vortex, the magnetization is rotated 
out of the easy plane, allowing the in-plane magnetization to vanish and $\phi$ to change discontinuously. The energy cost of creating vortices is therefore related in part to the strength of the easy-plane ansiotropy.  A non-zero uniaxial anisotropy energy is 
essential for the stability of the spin supercurrent \cite{PhysRevLett.87.187202}, as we emphasize again in the next 
subsection.  

As we have explained, the stability of supercurrent states in general superfluids can be understood in terms of the 
conservation of circulation, whether quantum or classical. However, we have not yet addressed the 
reason why the superfluids act like idea fluids, {\it i.e.} why viscosity (or dissipation) is absent. 
This issue will be discussed in the next subsection.

\subsection{Spin superfluidity and Bose-Einstein condensation}

The prototypical superfluid, liquid He$^4$, is also a Bose-Einstein condensate. Although the two concepts, superfluidity and BEC, are not equivalent, nor is one necessarily the consequence of the other, they are intimately related. 
In this subsection we will discuss the relationship between BEC and superfluidity, while at the same time making 
a comparison between BEC and easy-plane magnetism.

Briefly, to avoid repeating material presented in earlier Chapters, 
we define a BEC as a state of matter in which a macroscopic number of bosonic particles share the same single-particle wavefunction. 
For simplicity we assume here that {\it all} particles are in the same state.
One can then write the wavefunction of this state $\Psi$ as a direct product of the single particle states $\psi$:
\begin{eqnarray}
\Psi(\{{\bf r}_i\},t) = \prod_{i=1}^N \psi({\bf r}_i,t) \exp(-i \mu t/\hbar) 
\end{eqnarray}
where $\psi$ satisfies a mean-field Schr\"{o}dinger equation:
\begin{eqnarray}\label{eq:chen_GPequation}
i\frac{\partial \psi}{\partial t} = \left( - \frac{\hbar^2}{2m}\nabla^2 - \mu \right)\psi + \psi\int |\psi({\bf r}^\prime )|^2 N U({\bf r}-{\bf r}^\prime) d{\bf r}^\prime.
\end{eqnarray} 
Here $\mu$ is the chemical potential, and the last term is due to a weak interaction between particles.  
The factor $N$ in the last term reflects the fact that the effective interaction strength scales with the number of particles in the condensate. Equation \ref{eq:chen_GPequation} is called the (time-dependent) Gross-Pitaevskii (GP) equation\cite{gross1961,Pitaevskii1961} and is  
is central to the earliest and also the most widely used 
microscopic theory of BECs formed by weakly interacting bosonic particles. 
Below we will absorb a $\sqrt{N}$ factor into $\psi$. The integral of $|\psi|^2$ over space is then the total number of particles in the condensate. We can therefore regard $\psi$ as the order parameter of the condensate and the GP equations as an equation for order parameter dynamics. Below we emphasize its similarity to the Landau-Lifshitz equation for the order parameter dynamics of an easy-plane magnet.
The close relationship between these two equations is of course not coincidental.  

Assuming the BEC order parameter is a complex scalar function of position and time $\Psi = \sqrt{n({\bf r},t)} \, e^{i\phi({\bf r},t)}$, where $n$ is the density of the condensed particles, the time dependent GP equation can be rewritten as 
coupled equations for $n$ and $\phi$:
\begin{eqnarray}\label{eq:chen_GPnphi}
&&\hbar\dot{\phi} = \frac{\hbar^2}{2m}\left(\frac{1}{2n}\nabla^2n - \frac{1}{4n^2}|\nabla n|^2 \right) - \frac{\hbar^2}{2m}|\nabla \phi|^2 + \mu - U_0 n,\\\nonumber
&& \dot{n} = -\frac{\hbar}{m} \left(  n\nabla^2 \phi + \nabla n\cdot \nabla\phi \right),
\end{eqnarray}
where $U_0 ({\bf r})\equiv \int U({\bf r}-{\bf r^\prime}) d{\bf r}^\prime$. The 2nd equation has the form of continuity equation if the current is
\begin{eqnarray}
{\bf j} = n\frac{\hbar}{m}\nabla\phi \equiv n{\bf v}_s.
\end{eqnarray}
One can check that such a definition indeed agrees with that calculated from the standard formula ${\bf j} = -\frac{i\hbar}{2m}( \Psi^*\nabla\Psi-\Psi\nabla\Psi^*)$. Moreover, by taking
the time derivative of ${\bf v}_s$ and making use of the first equation in Eq.~\ref{eq:chen_GPnphi}, we obtain 
\begin{eqnarray}
\frac{\partial {\bf v}_s}{\partial t} + {\bf v}_s\cdot \nabla {\bf v}_s = -\nabla\left[ \frac{U_0n}{m}-\frac{\mu}{m}-\frac{\hbar^2}{2m^2}\left( \frac{1}{2n}\nabla^2n-\frac{1}{4n^2}|\nabla n|^2 \right)  \right],
\end{eqnarray}
which coincides with Euler's equation for an ideal fluid, Eq.~\ref{eq:chen_eulereqn}. Thus the condensate is an ideal fluid, conserves velocity circulation, and is irrotational ($\nabla \times {\bf v}_s=0$). Moreover, its angular momentum must be carried by quantized vortices as we discussed in the previous subsection. (These conclusions apply only when the BEC order parameter is a complex scalar function, and do not apply to spinor BECs.) The Landau-Lifshitz equations of easy-plane magnets, Eq.~\ref{eq:chen_conjeom},
correspond to the GP equations of BECs if we associate the term $\propto m_z$ in the first equation of the 
former with $\mu-U_0n$. (The counterpart of the kinetic energy term in Eq.~\ref{eq:chen_GPnphi} has been ignored in Eq.~\ref{eq:chen_conjeom} which considered the spatially constant order parameter case.) 
It is then interesting to ask if this means that an easy-plane ferromagnet can also be viewed as a BEC. Below we show that this is indeed the case.

Let us start from a single macrospin with angular momentum $s\hbar$, where $s$ is a real number much larger than $\frac{1}{2}$. Taking the $z$ direction to be the quantization axis, the raising and lowering operators for the $z$-spin are written as
\begin{eqnarray}
&&S_+=S_x+iS_y,\,\,S_+| s, s_z \rangle = \sqrt{s(s+1)-s_z(s_z+1)}\hbar | s, s_z+1 \rangle,\\\nonumber
&&S_-=S_x-iS_y,\,\,S_-| s, s_z \rangle = \sqrt{s(s+1)-s_z(s_z-1)}\hbar | s, s_z-1 \rangle 
\end{eqnarray}
where $| s, s_z \rangle$ is the eigenstate of $S_z$ with the eigenvalue $s_z\hbar$. 
Letting $S_{-}$ act repeatedly on eigenstates of $S_z$ generates a set of states $|n\rangle$, which are eigenstates of $S_z$ with
eigenvalues $(s-n)\hbar$. One can then define a set of bosonic creation and annihilation operators acting on these Fock states, 
$a^+$ and $a$, which decrease ($a^+$) or increase ($a$) the spin projection in $z$ direction by $\hbar$, i.e.,
\begin{eqnarray}
&&[a,a^+]=1,\\\nonumber
&&a|n\rangle =\sqrt{n}|n-1\rangle,\\\nonumber
&&a^+|n\rangle = \sqrt{n+1}|n+1\rangle.
\end{eqnarray}
$a^+$ and $a$ are related to $S_\pm$ and $S_z$ through the Holstein-Primakoff transformation \cite{hosteinprimakoff1940}
\begin{eqnarray}
&&S_+=\hbar\sqrt{2s-a^+a} \, a,\\\nonumber
&&S_-=\hbar\sqrt{2s-a^+a} \, a^+,\\\nonumber
&&S_z=\hbar(s-a^+a).
\end{eqnarray}   

We first ignore magnetic anisotropy altogether by assumming for the moment that the 
Hamiltonian commutes not only with the total spin component $S_z$, as it does in ideal easy-plane ferromagnets,
but also with $S_x$ and $S_y$. In this case all eigenstates occur in spin-multiplets and in the case of 
ferromagnets, the ground state multiplet has a macroscopic value of $s$, proportional to the size of the system.  
Now consider the ground state of an easy-plane ferromagnet, which should be 
an eigenstate of $S_x^2+S_y^2$. We define this state as $|XY\rangle$, which must have the property that 
\begin{eqnarray}
S_z^2|XY\rangle = \left[S^2-(S_x^2+S_y^2)\right]|XY\rangle = 0.
\end{eqnarray}
Therefore $|XY\rangle$ can be constructed using the bosonic operator $a^+$ acting on the vacuum--the eigenstate of $S_z$ with eigenvalue $s\hbar$:
\begin{eqnarray}
|XY\rangle = |s,s_z=0\rangle = \frac{1}{\sqrt{s!}}\left(a^+ e^{i\phi} \right)^s \, |0\rangle,
\end{eqnarray}
where $\phi$ is the azimuthal orientation angle of the macrospin. Therefore the ground state of an easy-plane ferromagnet can be viewed as a condensate of $N=M_{\rm tot}/(|g|\mu_B \hbar)$ $z$-spin Holstein-Primakoff bosons (magnons). 
When magnetic anisotropy is included, the ground state weakly mixes states with slightly different values of 
$s$, but this picture still applies. For an easy-spin magnet, the Landau-Lifshitz equation can
therefore be viewed as the counterpart of the GP equation for the $z$-spin magnon condensate. 
Quantum fluctuations in the local value of $S_z$ correspond to quantum fluctuations in boson density and,
quantum fluctuations in the azimuthal angle $\phi$ correspond to quantum fluctuations in the condensate phase.  
The correspondence between the $m_z$ term in the $\dot{\phi}$ equation in Eq.~\ref{eq:chen_conjeom} and $(\mu-U_0n)$ in Eq.~\ref{eq:chen_GPnphi} is also clear since both express the energy change associated with changing the particle number by one. We note that another way to understand the condensate nature of an easy-plane ferromagnet is through its analogy with the pseudospin description of superconductivity by Anderson\cite{anderson1958}, with electron-electron pairing in superconductivity replaced by electron-hole pairing in easy-plane ferromagnetism \cite{PhysRevLett.87.187202}.

It is now time to discuss the origin of vanishing viscosity in superfluids in relationship to analogous properties of 
easy-plane ferromagnets. First we discuss the analog of the Landau's criterion for superfluidity, namely that 
the system be in a metastable state that cannot relax to the ground state via elementary excitations, which we now briefly summarize.
(Vortex nucleation requires an unbounded energy and is not an elementary excitation.) A fluid flowing with velocity $\bf v$ has kinetic energy $E=\frac{1}{2}Mv^2$. Consider the possibility of energy dissipation through creation of an elementary excitation that has energy $\epsilon$ and momentum $\bf p$ in the reference frame moving with the fluid. One can find that in the rest frame the energy of the excitation is $\epsilon+{\bf p}\cdot {\bf v}$. The moving fluid is metastable (ignoring thermal excitations at finite temperature) only if all excitations have positive energy in the rest frame, {\it i.e.} only if the velocity of the fluid
\begin{eqnarray}
v < \min\left( \frac{\epsilon}{p} \right).
\end{eqnarray}
If the elementary excitations of the fluid have linear dispersion, this criterion can be satisfied below 
a critical velocity. Indeed, the elementary excitations in weakly interacting boson systems (as in superfluid He$^4$) are sound waves with linear dispersion as can be derived by linearizing the GP equation around its ground state solution $\Psi$.
Easy-plane ferromagnets are also superfluids in the same sense that their finite spin-current states can decay to the ground states only via vortex-nucleation processes and not via elementary excitations. As in a BEC, an easy-plane ferromagnet 
has linearly dispersing spin waves as elementary excitations. This result can be established by taking the 2nd order time derivative of $\phi$ and making use of the Landau-Lifshitz equation Eq.~\ref{eq:chen_conjeom} to obtain 
\begin{eqnarray}\label{eq:chen_ddotphi}
\ddot{\phi} = 4\gamma^2M_0^2AK\nabla^2\phi.
\end{eqnarray}
The spin-wave velocity 
\begin{eqnarray}\label{eq:chen_jcrit}
c = 2|\gamma|M_0\sqrt{AK}
\end{eqnarray}
is identical to the upper critical value of the spin supercurrent \cite{PhysRevLett.87.187202}. The linearly dispersive elementary excitations in both BEC and easy-plane ferromagnets are Goldstone modes related in the magnet case to spontaneous rotational symmetry breaking and in the BEC case to gauge symmetry breaking. Isotropic ferromagnets are not spin superfluids because 
their magnon dispersion is quadratic rather than linear at long wavelengths \cite{sonin_review}. Landau's criterion is, however, not a sufficient condition for superfluidity, since it says nothing about the topological stability of the metastable superfluid states. 

We end this section by noting that a discussion of superfluidity normally starts from the identification of a well defined current. In other words, from a continuity equation that can be written down for the physical quantity that is transported without dissipation and whose total number is conserved. This is not a problem with the mass superfluidity in BEC or the charge superfluidity in superconductors, since particle number is a good quantum number in both cases. However, no component of spin is ever really a good quantum number due to inevitable spin-orbit coupling and magneto-static interaction processes. The concept of spin currents has 
nevertheless been useful in spintronics, because spin is {\it nearly} conserved. The use of this concept does however sometimes lead to debate and confusion \cite{PhysRevLett.96.076604, sonin_review}, especially in cases where spin-orbit coupling plays a dominant role \cite{brataas_sot_review}. 
In fact, the easy-plane anisotropy required for a finite critical current in our spin superfluid obviously requires spin-orbit coupling. If there is no other anisotropy the $z$ component of total spin is still a good quantum number, which means the $z$-spin supercurrent is well defined. 
In reality, however, there is always some anisotropy in the easy plane. The fact that $S_z$ is not conserved leads to both dissipative and reactive effects which must both be taken into account in analyzing spin-transport phenomena. When we invoke the concept of spin-superfluidity we have in mind the metastability of magnetic configurations that carry spin-currents through a system collectively through the magnetic condensate, and not via non-equilibrium magnon or electron quasiparticles. In the next section we will discuss realistic situations and show how the concept of spin superfluidity is useful even though $S_z$ is not a good quantum number.  

\section{Dynamics of spin superfluids with spin injection}   

The central idea of spintronics is that spin can be used instead of or as a complement to 
charge to carry information through circuits and to store information. When spin-orbit coupling is negligible, total spin is a good quantum 
number. One can then define the spin current by multiplying spin with the probability current operator $j$, 
for example $S_zj$ is the spin current operator for the $\hat{z}$-spin projection (see below). 
One therefore needs to trace over the spin degree of freedom to get the expectation value of the spin current. It is possible to have a spin current that is not accompanied by net charge transport, a pure spin current, when the charges carried by states with opposite spins cancel. Since spins couple to lattice vibrations much more weakly than charges, the Joule heating problem associated with 
electronics-based circuits could be mitigated if charge and spin transport could be decoupled.  

In the absence of spin-orbit coupling and magnetostatic interactions 
\begin{eqnarray}
\dot{\bf S} = \frac{i}{\hbar}[H,{\bf S}] = 0;
\end{eqnarray}
in other words spin is a good quantum number. For an individual independent electron
\begin{eqnarray}\label{eq:chen_dotSwoSO}
\frac{\partial \langle {\bf S}({\bf r},t) \rangle}{\partial t} =  - \langle {\bf S} \otimes \left(\nabla\cdot {\bf j} \right ) \rangle \equiv -\nabla\cdot \langle \hat {j}_S \rangle,
\end{eqnarray}
where $\bf j$ is the usual probability current operator in quantum mechanics, and 
$\hat{j}_S$ is the spin current operator which is a rank 2 tensor. 
When $\bf S$ is not a good quantum number, 
\begin{eqnarray}
\dot{\bf S} = \frac{i}{\hbar}[H,{\bf S}] \equiv \Pi \neq 0,
\end{eqnarray}
where $\Pi$ is the spin torque operator, and 
\begin{eqnarray}\label{eq:chen_dotSwSO}
\frac{\partial \langle {\bf S} \rangle({\bf r},t)}{\partial t} = \frac{i}{\hbar} \left[ (H\psi)^\dag {\bf S}\psi-\psi^\dag {\bf S}(H\psi) \right] + \psi^\dag \Pi \psi.
\end{eqnarray}
One cannot isolate a current from the right hand side of Eq.~\ref{eq:chen_dotSwSO} in any unambiguous way. Even in the case that the first term on the right hand side of Eq.~\ref{eq:chen_dotSwSO} can be approximately identified as the 
divergence of the spin current defined in Eq.~\ref{eq:chen_dotSwoSO}, the torque term can still change the spin density locally even with a uniform steady effective spin current. If one insists on maintaining the 
same definition of spin current, this torque term accounts for 
additional sources and sinks of spins.

It should be acknowledged that spin-currents are in fact normally accompanied by dissipation. We distinguish two classes of mechanisms.
(i) Dissipation associated with diffusive motion of magnon or electron quasiparticles: Quasiparticle scattering tends to relax the quasiparticles toward a state that is at rest with respect to the lattice, and in the process to transfer energy to phonons or magnons. In this case the dissipation can be described by classical Boltzmann theory. There is little difference, particularly if spin is carried by electronic quasiparticles, between the dissipation associated with quasiparticle charge currents and spin currents.
(ii) Dissipation due to relaxation of the magnetic condensate toward its minimum energy configuration. This type of dissipation is captured by the Gilbert damping terms which appear in the Landau-Lifshitz equations for collective dynamics. No analogous terms appear in the GP equations for an equilibrium BEC.  Similar terms 
do appear however in phenomenological descriptions of magnon condensates, which are always non-equilibrium steady states
that are not true thermal equilibrium.  
By exploiting spin supercurrents in an easy-plane ferromagnet one can largely get rid of the dissipation due to the first mechanism.
If $S_z$ is conserved, the spin supercurrent is well defined and one can use the easy-plane ferromagnet as a 
dissipationless link to efficiently transport spin between remote spintronics devices. 

To understand the role of Gilbert damping and magnetic anisotropy within the easy plane we need to study the dynamics of 
spin superfluids subject to injection or extraction of normal spin currents, which is discussed in the next subsection. 
The spin spiral states of Fig.~\ref{fig:chen_fig1} which carried a spin supercurrent in the ideal case, are slightly distorted by weak in-plane
magnetic anisotropy, but their metastability is largely unaffected. In Sec.~\ref{sec:chen_NSN} we discuss a possible spintronic
device based on easy-plane ferromagnets that is conceptually similar to a N-S-N circuit containing normal metal leads connected to a superconducting wire.  

\subsection{Dynamics of spin superfluids with spin injection}
In this subsection we describe the basic ideas needed to understand spin supercurrents in a finite easy-plane 
ferromagnet coupled to external sources/drains of quasiparticle spin. 
The spintronics toolkit contains a variety of possible sources of spin-currents with spin polarization perpendicular to the easy plane, including ones based on the spin Hall effect, ferromagnetic resonance, or electron tunneling from perpendicular anisotropy magnetic films. 
Note that electrical generation of spin-currents always requires a charge bias potential. A normal spin current in an easy-plane ferromagnet can be supported by electronic quasiparticles only close to the current source. Assuming that $S_z$ is a good quantum number for now, the continuity equation for $S_z$ in this 
boundary layer guarantees that this current 
will be converted into a collective spin-supercurrent:  
\begin{eqnarray}
{\bf j}_{nz}= 2\gamma A M_0 \nabla\phi\big|_B,
\end{eqnarray}
where ${\bf j}_{nz}$ is the $z$-spin current injected from the source, and the 
subscript $B$ indicates
that the spatial derivative of the azimuthal magnetization orientation $\phi$ should be evaluated at a position close to the source or drain. 

By eliminating $m_z$ in the Landau-Lifshitz equation Eq.~\ref{eq:chen_conjeom}, the dynamics of $\phi$ in the bulk of the easy-plane ferromagnet is described by Eq.~\ref{eq:chen_ddotphi}. For simplicity we consider a one dimensional problem.
In the steady state $\phi(x,t)=\phi(x)-\omega t$, and $\phi(x)$ is the solution of
\begin{eqnarray}
&&\partial_x^2\phi=0,\\\nonumber
&&{\bf j}_{nz}= 2\gamma A M_0 \nabla\phi\big|_B,
\end{eqnarray}
where the boundary condition must be satisfied at both ends of the 1D system. These conditions yield 
\begin{eqnarray}\label{eq:chen_steadystate0}
&&\phi(x,t)=\frac{j_{nz}}{2\gamma AM_0}x-\omega t,\\\nonumber
&&{\bf j}_{nz,L}={\bf j}_{nz,R}.
\end{eqnarray}
The easy-plane ferromagnet is driven to a spiral state with wave vector 
\begin{equation}
q=\frac{j_{nz}}{2\gamma AM_0},
\end{equation}
and the net spin current injected into the system must be zero or the system would not be able to find a steady state. $j_{nz}$ also has to be smaller than the critical value given in Eq.~\ref{eq:chen_jcrit} in order for the supercurrent state to be sustained.

To understand the significance of the spin-precession frequency we transform the spin part of the system into a rotating frame synchronized with the precession of the order parameter. The unitary operator which achieves this transformation is 
\begin{eqnarray}
U=e^{-i\frac{\omega t}{2}\sigma_z}.
\end{eqnarray}
In the mean-field Hamiltonian of the easy-plane ferromagnet the time-dependent order parameter leads to a term proportional to $\cos(qx-\omega t)\sigma_x+ \sin(qx-\omega t)\sigma_y$. Applying the unitary transformation on this operator yields
\begin{eqnarray}
U\left[\cos(qx-\omega t)\sigma_x+ \sin(qx-\omega t)\sigma_y \right]U^\dag = \cos(qx)\sigma_x+ \sin(qx)\sigma_y,
\end{eqnarray}
{\it i.e.} the precession is removed. The tradeoff is that the Hamiltonian acquires a spin-dependent chemical potential shift, which can be seen from the modification to the time-evolution operator
\begin{eqnarray}\label{eq:chen_dotphiU}
|\psi(t)\rangle_R = U |\psi(t)\rangle = e^{-i\frac{\omega t}{2}\sigma_z} e^{-i\frac{Ht}{\hbar}}|\psi(0)\rangle =  e^{-i\frac{t}{\hbar}(H+\frac{\hbar\omega}{2}\sigma_z)}|\psi(0)\rangle_R.
\end{eqnarray}
Note that the last equality requires $S_z$ to be conserved. This equivalence between dynamics and spin-dependent chemical potential is well known in spintronics where is it responsible for spin-pumping \cite{TserkovnyakRevModPhys.77.1375} and spin 
electromotive forces \cite{emf-Berger,PhysRevLett.102.067201}. 

\begin{figure}
\includegraphics[width=6.5cm]{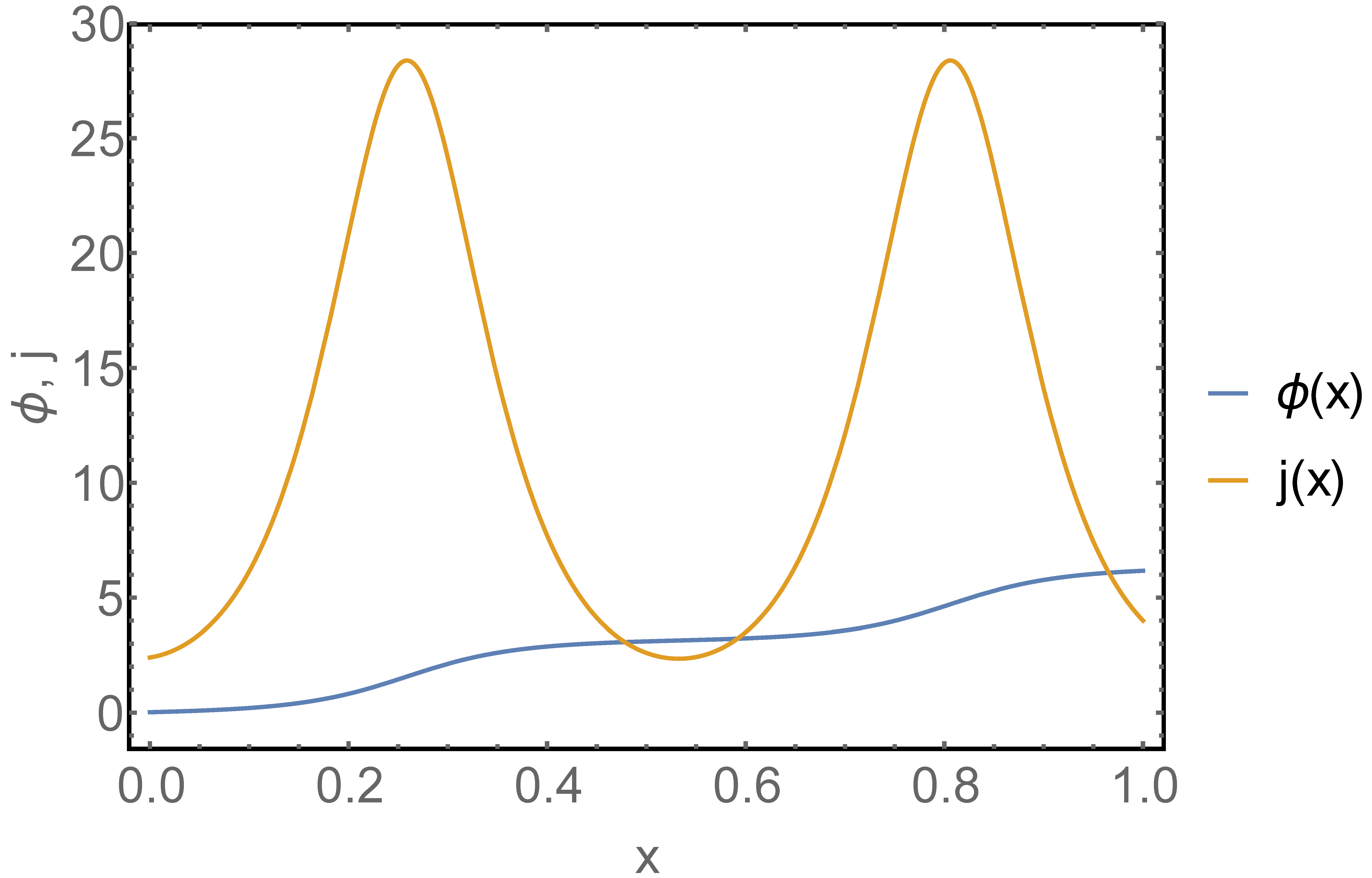} 
\caption{Distorted supercurrent spiral in a finite 1D system with spin injection at the 
sample ends and uniaxial easy-axis anisotropy along $\hat{x}$. }
\label{fig:chen_fig2}
\end{figure}

We now consider the effect of adding in-plane uniaxial anisotropy along the $x$ direction to the magnet's energy functional:  
\begin{eqnarray}
-K^\prime M_x^2=-\frac{1}{2} K M_0^2\cos(2\phi)+{\rm const},
\end{eqnarray}
where the constant term can be ignored. The discussion below can be easily generalized to other forms of anisotropy. 
A Hamiltonian contribution which gives rise to this anisotropy obviously does not commute with the $z$ component of spin in the microscopic Hamiltonian. As a result the $z-$spin current is rigorously speaking not a well defined quantity. Nevertheless as we have discussed earlier in the approximation that the spin density varies slowly in space we can still use the spin current language and separate the contribution to $\dot{m}_z$ into a current term and a torque term. This can be seen in the Landau-Lifshitz equations modified by this anisotropy:
\begin{eqnarray}\label{eq:chen_LLaniX}
&&\dot{\phi}=2\gamma KM_0 m_z,\\\nonumber
&&\dot{m}_z = 2\gamma A M_0\nabla^2\phi - \gamma K^\prime M_0\sin(2\phi),
\end{eqnarray} 
where we have assumed $K^\prime \ll K$ and on this basis ignored its modification to the $\dot{\phi}$ equation. The 2nd term on the right hand side of the $\dot{m}_z$ equation is the extra torque from anisotropy within the easy-plane. Eliminating $m_z$ from Eq.~\ref{eq:chen_LLaniX} we obtain the sine-Gordon equation
\begin{eqnarray}\label{eq:chen_ddotphianiX}
\ddot{\phi} - c^2\left[ \nabla^2\phi - \frac{\sin(2\phi)}{l^2} \right] = 0,
\end{eqnarray}
where $c$ is given in Eq.~\ref{eq:chen_ddotphi}, and $l=\sqrt{2A/K^\prime}$. The simplest
time-independent solution of Eq.~\ref{eq:chen_ddotphianiX} contains a single soliton (domain wall):  
\begin{eqnarray}
\phi(x) = 2\arctan\left[ \exp\left( \sqrt{2}\frac{x - a}{l}\right)\right],
\end{eqnarray}
where $a$ is the arbitrary soliton position. The homogeneous spiral state in the absence of the easy-axis anisotropy is thus not a stable state of the system; for any given phase gradient the system can lower its energy by locally rotating the in-plane polarization toward its easy axis, thereby distorting the simple spiral state. The collective spin supercurrent is nonuniform in space, with its divergence matching the rate of transverse spin creation or annihilation by the torque from the in-plane anisotropy.  
It is often still possible, however, to find metastable distorted spiral states which satisfy the boundary conditions imposed by spin-currents injected or absorbed at sample boundaries by solving a boundary value problem with Neumann boundary conditions:
\begin{eqnarray}\label{eq:chen_PDEaniX}
&&\partial_x^2\phi - \frac{\sin(2\phi)}{l^2}=0,\\\nonumber
&&{\bf j}_{nz}\big|_{L,R}= 2\gamma A M_0 \partial_x\phi\big|_{L,R}.
\end{eqnarray}
Strictly speaking the boundary conditions should include a spin torque term due to the easy-axis anisotropy at the boundary. 
However, since the torque contribution is an integral over the volume of the boundary layer, we can always ignore this term provided that the boundary layer is thin enough. An example of the solution of Eq.~\ref{eq:chen_PDEaniX} is shown in Fig.~\ref{fig:chen_fig2}.

Since static solutions balance spatial variation in spin currents against the in-plane anisotropy torque, it is clear that when the net current injection exceeds a value determined by the easy-axis anisotropy, a static solution may not be found. An estimate of the critical current imbalance can be made by assuming the stiffness $A$ is very large, so that both the wavelength of the spiral (Eq.~\ref{eq:chen_steadystate0}) and the width of the domain wall $l$ greatly exceed the system size. In this macrospin limit Eq.~\ref{eq:chen_PDEaniX} reduces to
\begin{eqnarray}
-\gamma K^\prime M_0 V \sin(2\phi) = {\bf I}_L+ {\bf I}_R,
\end{eqnarray}  
where $V$ is the volume of the easy-plane ferromagnet and ${\bf I}_{L,R}$ are the normal spin currents injected. In this limit the critical current imbalance is 
\begin{eqnarray}\label{eq:chen_Inetcrit}
\max(| {\bf I}_L+ {\bf I}_R |) = |\gamma| K^\prime M_0 V\equiv I_c.
\end{eqnarray}
A discussion of the opposite limit that $l\ll L$ where $L$ is the length of a long easy-plane ferromagnet can be found in \cite{chenPhysRevB.90.220401}. Note that static solutions are always available when the spin-current injected at one end of the 
sample is equal to the spin-current removed at the other end of the sample.  

The order parameter is not static when there are no metastable magnetic configurations that satisfy spin-injection boundary conditions. Under such circumstances it is necessary to consider its damping. Collective magnetization dynamics, including damping, 
is accurately described by the Landau-Lifshitz-Gilbert (LLG) equation when magnetic order is well developed:
\begin{eqnarray}\label{eq:chen_LLG}
\frac{d{\bf M}}{dt} = -\gamma {\bf M} \times \frac{\delta \mathcal{F}}{\delta {\bf M}} + \frac{\alpha}{M_0} {\bf M}\times \frac{d{\bf M}}{dt},
\end{eqnarray}
where $\alpha$ is the Gilbert damping parameter. Taking the in-plane easy axis anisotropy into account, the LLG equation in terms of $\phi$ and $m_z$ is  
\begin{eqnarray}\label{eq:chen_LLGaniX}
&&\dot{\phi}=2\gamma KM_0 m_z -\alpha \, \dot{m}_z,\\\nonumber
&&\dot{m}_z = 2\gamma A M_0\nabla^2\phi - \gamma K^\prime M_0\sin(2\phi) + \alpha \dot{\phi}.
\end{eqnarray}

Solving Eq.~\ref{eq:chen_LLGaniX} in a finite system is challenging in general. Here we only consider the macrospin limit and assume a steady state in which $\dot{\phi}$ is spatially constant. For large easy-plane anisotropy this means that $\dot{m}_z=0$ according to the $\dot{\phi}$ equation in Eq.~\ref{eq:chen_LLGaniX}. We thus arrive at a single equation for $\phi$:
\begin{eqnarray}\label{eq:chen_pendulum}
-\gamma K^\prime M_0 V \sin(2\phi) + \alpha V\dot{\phi} = {\bf I}_{\rm net}.
\end{eqnarray} 
For $|I_{\rm net}|\gg I_c$. where ${\bf I}_{\rm net} = {\bf I}_{L}+{\bf I}_{R}$, the solution is approximated by $\phi(t)=\phi_0 + (I_{\rm net}/ \alpha V) \, t$. When $|I_{\rm net}|\sim I_c$, $\phi(t)$ has an additional oscillatory contribution. ({\it cf.} Fig. 1b in \cite{chenPhysRevB.90.220401}). 

An important consequence of having both in-plane anisotropy and Gilbert damping in the easy-plane ferromagnet is that it is
possible to drive the easy-plane ferromagnet across the transition between two very different spin-transport regimes. Specifically, recall that the precession of the in-plane magnetization is equivalent to a spin-dependent chemical potential shift $\delta\mu=-\frac{\hbar\dot{\phi}}{2}\sigma_z$ (Eq.~\ref{eq:chen_dotphiU}). When the magnetization is static $\delta \mu =0$ even for finite $I_{\rm net}<I_c$ because of the easy-axis anisotropy within the easy
plane, whereas in the steady precessing state $\delta \mu\approx- (\hbar I_{\rm net}/2\alpha V) \, \sigma_z$, which can be very large when damping is small. The current dependence of the spin voltage in the system is thus highly nonlinear. In the next subsection we will study this behavior in more detail and explore its potential use.

\subsection{Device based on a N-S-N junction}\label{sec:chen_NSN}

\begin{figure}
\includegraphics[width=5.5cm]{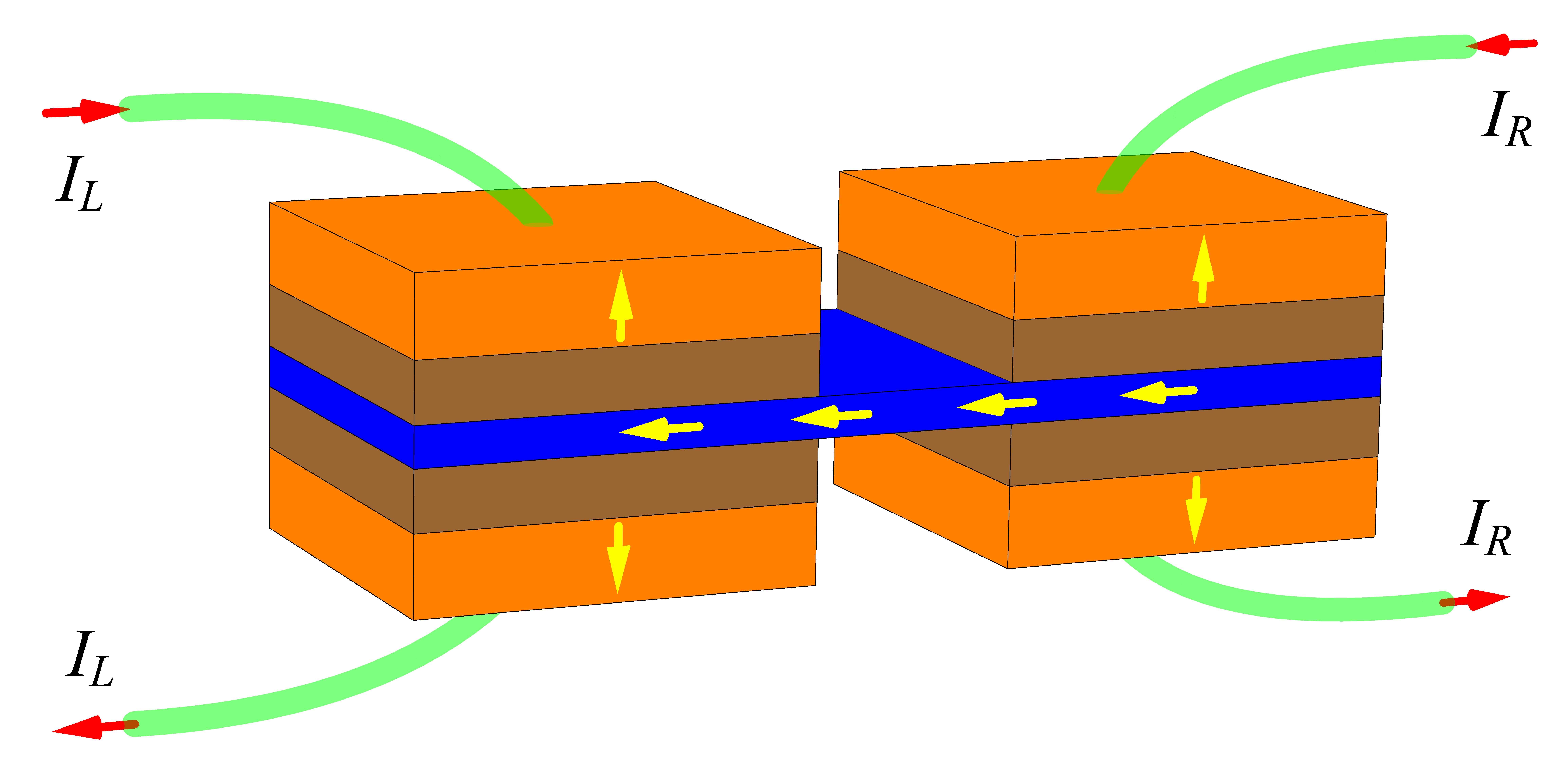} 
\caption{A schematic illustration of the bi-stack magnetic transistor concept.}
\label{fig:chen_fig3}
\end{figure}

In this subsection we study a structure formed by an easy-plane ferromagnet sandwiched between two perpendicular anisotropy ferromagnetic tunnel junctions, as schematically illustrated in Fig.~\ref{fig:chen_fig3}. A ferromagnetic tunnel junction is formed by two easy-axis ferromagnets with opposite magnetizations, separated by dielectrics. When a tunneling current is established in the junction, $z$-spin conservation dictates that there must be pure spin currents injected into the easy-plane magnetic system at the position of the tunnel junction stack. These spin currents can be carried collectively from one stack to the other, even when the easy-plane system is not metallic. Because the quasiparticle spin currents in the ferromagnetic tunnel junctions are converted into spin supercurrents in the easy-plane ferromagnet, this geometry provides a magnetic analog of a N-S-N circuit.

The spin N-S-N junction can also be described using a microscopic model suitable for nonequilibrium Green's function calculations, which we briefly introduce here. The left and right metal stacks can be represented by nearest neighbor tight-binding models with 
no spin-orbit coupling and a difference between the up and down spin chemical potentials. To model the easy-plane magnet, we add to the tight binding model a mean-field on-site anisotropic interaction
\begin{eqnarray}\label{eq:chen_HaniMF}
H_A = \sum_i \sum_{\alpha=x,y,z} U_\alpha S_{i\alpha}\langle S_{i\alpha}\rangle,
\end{eqnarray}
and set $U_x=U_y < U_z$ to account for the easy-plane or hard-axis anisotropy. $H_A$ is also responsible for spontaneous magnetic ordering. This microscopic model complements the macroscopic Landau-Lifshitz description in the previous subsection by providing information on, {\it e.g.} the dependence of the magnitude of the in-plane magnetization on the potential biases in the leads, the decay length of normal spin currents injected into the easy-plane ferromagnet, and on the difference in behavior between insulating and metallic easy-plane ferromagnets. The model can also be used to study spin superfluidity in antiferromagnets since the on-site interaction Eq.~\ref{eq:chen_HaniMF} is more likely to lead to antiferromagnetic ground state in equilibrium.

A benefit of using the ferromagnetic tunnel junctions to inject spin currents into the easy-plane ferromagnet is that the size of the spin current is directly determined by the electric voltages applied across the junctions. The magnetization dynamics of the easy-plane ferromagnet influences transport through the perpendicular magnetic anisotropy stacks through the effect we mentioned at the end of the last subsection. By contacting two ferromagnetic tunnel junctions to the same easy-plane ferromagnet, it is possible to realize highly nonlinear and nonlocal current-voltage characteristics, particularly when the easy-plane ferromagnet is driven across the transition between static and precessing states. Such a device has potential application as a field-effect transistor. 
Similar proposals have been made using other condensed matter systems, {\it e.g.} spatially indirect exciton condensates \cite{HongkiBISFETPhysRevB.78.121401,BISFET4729616}.

To continue the analysis we stay with the large easy-plane anisotropy and macrospin limit, for which Eq.~\ref{eq:chen_pendulum} applies. It is however more relevant to use electric voltages across the ferromagnetic tunnel junctions instead of spin currents to 
characterize circuit characteristics. From the continuity equation of $z$-spin in the region of the ferromagnetic tunnel junctions it follows that the spin current is proportional to the tunneling charge current, {\it i.e.} that 
\begin{eqnarray}\label{eq:chen_ILRstatic}
I_{L,R} = \frac{F_{L,R}}{e}g_{L,R}U_{L,R},
\end{eqnarray} 
where $g_{L,R}$ is the tunnel conductance, $U_{L,R}$ is the bias voltage across the tunnel junction, and $F_{L,R} \leq 1$ is a system dependent parameter characterizing the conversion efficiency between charge (number) current and spin ($m_z$) current. 
When the in-plane magnetization of the easy-plane ferromagnet starts to precess, $U_{L,R}$ will be shifted by $-\hbar\dot{\phi}/e$ in the rotating frame of the easy-plane ferromagnet. It follows that in this case 
\begin{eqnarray}\label{eq:chen_ILRprecessing}
I_{L,R} = \frac{F_{L,R}}{e} g_{L,R} \left(U_{L,R}-\frac{\hbar\dot{\phi}}{e}\right).
\end{eqnarray} 
Eq.~\ref{eq:chen_pendulum} then becomes
\begin{eqnarray}\label{eq:chen_FETprecessing}
I_c \sin(2\phi) + g_i\frac{\hbar\dot{\phi}}{e^2} =
\frac{F_{L}}{e} g_{L} \left(U_{L}-\frac{\hbar\dot{\phi}}{e}\right) + \frac{F_{R}}{e} g_{R} \left(U_{R}-\frac{\hbar\dot{\phi}}{e}\right),
\end{eqnarray}
where $I_c$ is given in Eq.~\ref{eq:chen_Inetcrit}, and 
\begin{eqnarray}\label{eq:chen_gi}
g_i \equiv \alpha V e^2/\hbar
\end{eqnarray}
characterizes the Gilbert damping induced dissipation. When one increases $U_L+U_R$ so that $I_{\rm net}$ 
greatly exceeds $I_c$, the first term in Eq.~\ref{eq:chen_FETprecessing} vanishes after time averaging. 
Combining Eqs.~\ref{eq:chen_FETprecessing} and \ref{eq:chen_ILRprecessing} yields 
\begin{eqnarray}\label{eq:chen_IUprecessing}
I_L^e=\frac{g_i+F_R g_R}{g_i+F_R g_R+F_L g_L} g_L U_L - \frac{F_R g_L}{g_i+F_R g_R+F_L g_L} g_R U_R,
\end{eqnarray}
where $I_L^e$ means the tunneling electron current at the left tunnel junction. A similar equation for $I_R^e$ can be obtained by interchanging $L$ and $R$ labels. A non-local correlation between the \textit{charge} currents and voltages at the two tunnel junctions is thus established through the easy-plane ferromagnet, even when no charge paths connect the tunnel junction stacks.  

The static and precessing regimes discussed above are partly analogous to the DC and AC Josephson effects in superconductors \cite{Josephson1962251,tinkham_superconductivity}. The essence of the DC Josephson effect is that when the order parameter is static, the supercurrent is dependent on the position dependence of the condensate phase. A current can flow even when the voltage drop measured along the superconductor vanishes. In the AC Josephson effect the order parameter phase is linearly increasing on time with a constant rate of change proportional to the voltage applied across the superconductor. 

\begin{figure}
\includegraphics[width=6.5cm]{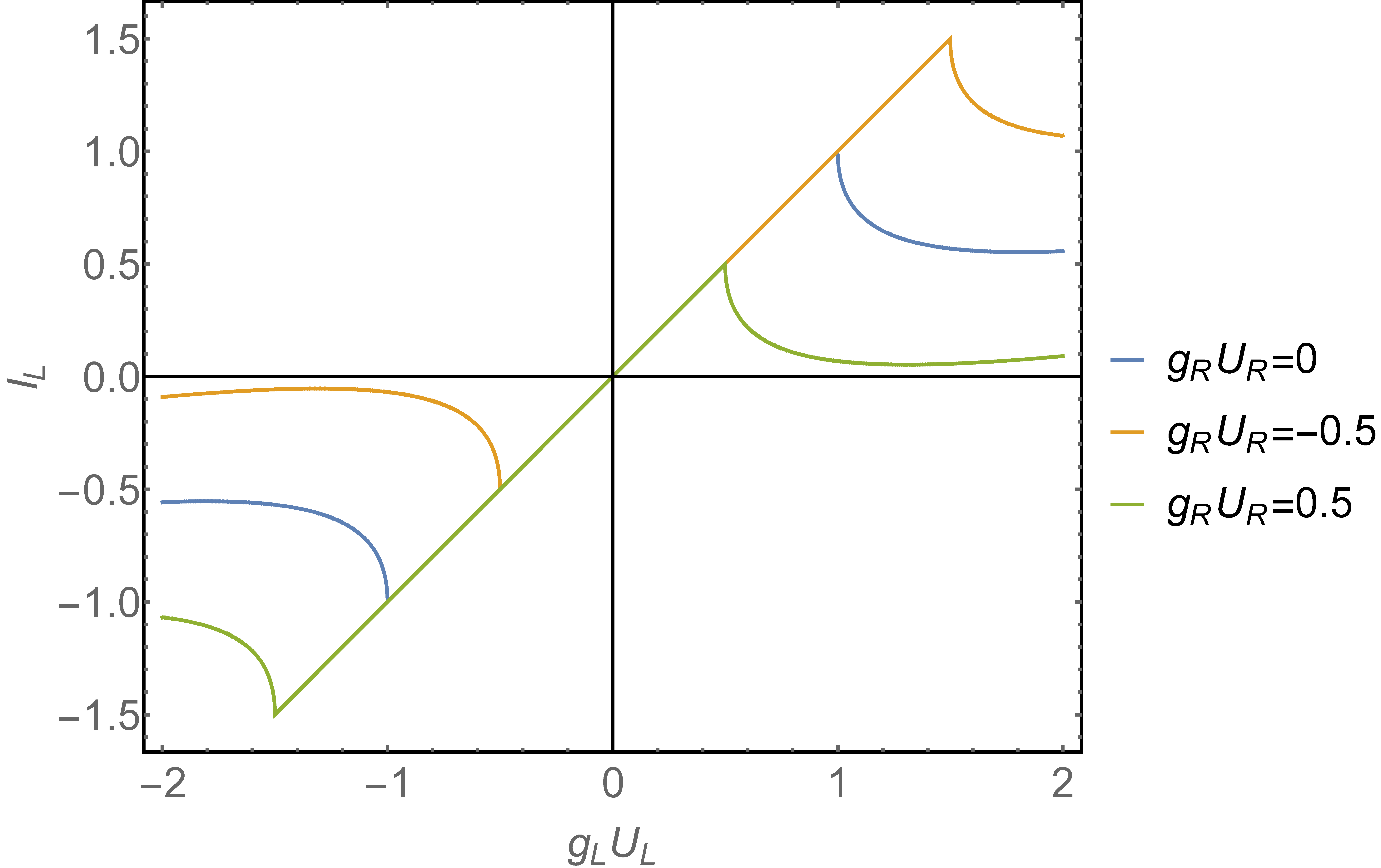} 
\caption{DC Current-voltage relationship of the device in Fig.~\ref{fig:chen_fig3}.}
\label{fig:chen_fig4}
\end{figure}

Comparing Eq.~\ref{eq:chen_IUprecessing} to the static case in which $I_L^e$ is simply equal to $g_L U_L$, we find that the effective conductance (with $U_R$ fixed) is reduced by a factor of 
\begin{equation} 
r = \frac{g_i+F_R g_R}{g_i+F_R g_R+F_L g_L}.
\end{equation}
The conductance reduction factor $r$ reflects the property that when the critical current is 
exceeded, electrons can no longer flip their spins by scattering off the easy-plane magnetic condensate 
and must instead take advantage of the incoherent process that contribute to Gilbert damping in order
to make their way through the stack. $r$ can in principle be much smaller than 1 if $g_i+F_R g_R \ll F_L g_L$, 
providing two states distinguished by very different DC resistances. Note that $g_i$ is proportional to nano-particle volume whereas $g_R$ and $g_L$ are proportional to stack areas, so that large conductance reduction can be achieved only in high quality thin film nanomagnets. Moreover, the transition between these two states can be controlled by $U_R$ since it
is determined by $I_{\rm net}$ (Eq.~\ref{eq:chen_Inetcrit}) or $U_L+U_R$. The device behaves very much like a field effect transistor (FET) and can be used as a switch. The typical current-voltage characteristics of the device is shown in Fig.~\ref{fig:chen_fig4}. 
We note that when $|I_{\rm net}|$ increases slightly above $I_c$ from below the charge current will have a large AC component while the DC component has a sudden drop.

The performance of a switch is evaluated based mainly on the following three considerations: the on/off ratio, the switch voltage (voltage around which the switching occurs), and the stability of the switching behavior against thermal fluctuations. We already see that small Gilbert damping $\alpha V$ (cf. Eq.~\ref{eq:chen_gi}) and effective spin current injection (large $F_{L,R} g_{L,R}$) are necessary for large on/off ratio. Permalloy is likely a suitable candidate for the easy-plane ferromagnetic junction because of its weak damping, and also because of the small anisotropy which the switch voltage is proportional to (cf. Eq.~\ref{eq:chen_Inetcrit}). Since $g_i\propto V$ it is ideal if the cross-sectional area is dominated by ferromagnetic stacks rather than by the easy-plane link part. The thermal stability of the switch is determined by the energy barrier between the static and the precessing states \cite{PhysRevLett.92.088302, PhysRevLett.93.166603, PhysRevLett.22.1364}, which is the in-plane anisotropy energy $\sim K^\prime M_0^2 V$. Therefore for the device to be operational, the minimal voltage difference between the on and the off states is $\delta U \sim (k_B T/K^\prime V)  \times (e^2 I_c)/(F_L g_L)$ should satisfy 
\begin{eqnarray}
\frac{\delta U}{k_B T} \sim \frac{1}{M_0 F_L g_L}\ll 1, 
\end{eqnarray}
where $M_0$ and $g_L$ are in units of $\mu_B$ and $e^2/\hbar$, respectively. This relation means that because of the collective nature of the switching phenomena, fundamental limits on conventional devices based on single electron behavior, can be circumvented.

\section{Discussion and conclusions}

In this chapter we have explained that ideal easy plane magnets can be viewed as equilibrium magnon Bose-Einstein condensates.
Magnon condensation in equilibrium differs qualitatively from condensation in systems with steady-state non-equilibrium populations of magnons, even when these partially thermalize. Just as Bose-Einstein condensation occurs in systems with conserved particle number, ideal equilibrium magnon condensation occurs in easy-plane magnetic systems in which the perpendicular $\hat{z}$ component 
of spin is a good quantum number. In these ideal systems a spiral magnetization configuration is metastable and carries a spin current without dissipation.  

In realistic cases, no component of spin is conserved. The concept of spin currents is nevertheless useful in both paramagnetic and ferromagnetic metals, even though it is necessary to be cautious in using the spin current concept which is sometimes 
ambiguous. This is also true for easy plane magnetic systems regarded as spin superfluids. The spin-current contribution to collective spin-dynamics which is readily identified in ideal systems is still present in the Landau-Liftshitz equations, which are the magnetic analog of the GP equations. There are still metastable magnetization configurations which carry spin-currents without dissipation, although the spin-current is not spatially constant because of the influence of torques associated with weak anisotropy within the easy plane. The dissipationless spin supercurrents are responsible for non-local relationships between the I-V characteristics of remote magnetic circuits which are coupled only by 
interacting with the same magnetic condensate.     

It is instructive to compare the properties of a system in which a bias voltage is applied across a superconductor by normal metal leads connected to a power supply, a N-S-N system, with the properties of a system in which an easy-plane magnet is connected to perpendicular magnetic anisotropy leads. In the superconductor case the two normal metal leads do not normally have spin accumulation,
{\it i.e.} they don't have well defined chemical potentials for $\uparrow$ and $\downarrow$ spins that are different. In the magnetic case, spin-accumulation is a common mechanism for 
the creation of spin-currents. A spin accumulation can be established either by illumination at a magnetic resonance frequency or by applying a charge bias voltage. 

The current which flows across a N-S interface is proportional to the chemical potential difference between the lead and the superconductor. The chemical potential 
of the superconductor is proportional to the time derivative of the condensate phase. In the steady state its value is adjusted so that the current flowing into the superconductor across one N-S interface is exactly equal to the current flowing out of the superconductor across the other N-S interface. In the macrospin limit the corresponding equation for the magnetic system is (cf. Eq.~\ref{eq:chen_FETprecessing}) 
\begin{equation} 
\hbar \dot{\phi} = \frac{ g_L F_L e U_L + g_R F_R e U_R }{g_i + g_L F_L + g_R F_R}.  
\end{equation} 
The left hand side of this equation is effectively a chemical potential for magnons, measured from the ground state 
chemical potential. If $g_i$ in this equation is set to zero, the magnon chemical potential will adjust to 
guarantee that the spin-current injected at one end is emitted from the following end. The total spin-current which 
flows through the system will then depend only on the spin-accumulation difference between one end of the 
magnet and the other. In spintronics language $\hbar \dot{\phi}$ is viewed as generating a spin-pumping contribution to the spin-currents at each end of the system. The properties of the N-S-N junction and the easy-plane magnetic system are therefore quite similar when $g_i$ is smaller than the electrode conductances. 

There is another route which allows spin-supercurrent 
behavior to be revealed. In the N-S-N circuit, only the chemical potential difference between the two N electrodes influences transport. In the magnetic case we have the ability to separately control the spin accumulations $U_L$ and $U_R$ and can for example choose their values such that the total injected current is below its critical value even when
the individual injected currents are large in value. 
In this case the large spin currents injected at one contact do not excite magnetization dynamics only because
of the large compensated spin-supercurrent injected at the other contact.  The large 
spin-supercurrent is carried along the sample without 
dissipation.    

In conclusion we point out that a number of considerations that are known to be important have not been extensively discussed 
in this brief chapter, and in some cases are only now being addressed in the literature.  
 Among these we mention in particular the role of long-range magnetostatic interactions, which are a 
 serious complication in samples that are beyond the macrospin-limit in size,  and the possibility of using easy-plane 
antiferromagnetic materials \cite{AFMsuperPhysRevB.90.094408} instead of ferromagnetic materials.
In ferromagnets magnetostatic interactions tend to destabilize the homogeneous magnetic configurations from 
which the spiral configurations arise, in favor of configurations containing domains with different orientations.
This problem is interesting but perhaps mainly academic since magnetostatic interactions are less important in smaller systems, and the largest interest is in exploiting spin superfluid properties in nanoscale spintronic devices. 
Most of the observations made in this chapter apply equally well to ferromagnets and antiferromagnets, which have the advantages that magnetostatic interactions are absent and that dynamics are faster -- possibly enabling spintronic devices that can be switched very rapidly.

\noindent\textbf{Acknowledgement}
This work was supported as part of the SHINES, an Energy Frontier Research Center funded by the U.S. Department of Energy, Office of Science, Basic Energy Sciences under Award \# SC0012670.

\bibliographystyle{cambridgeauthordate}

\end{document}